\documentclass{ws-procs975x65}
\usepackage{hyperref}

\def\beq{\begin{equation}}
\def\eeq{\end{equation}}

\def\de{\partial}
\def\lef{\langle}
\def\re{\rangle}

\begin{document}

\title{On the soft limit of tree-level \\ string amplitudes}
\author{Massimo Bianchi and Andrea L. Guerrieri$^*$}

\address{Dipartimento di Fisica and INFN, Università di Roma ``Tor Vergata'',\\
Via della Ricerca Scientifica, 00133 Roma, Italy\\
$^*$E-mail: andrea.guerrieri@roma2.infn.it}

\begin{abstract}
We study the soft behavior of string scattering amplitudes at three level 
with massless and massive external insertions, relying on different techniques 
to compute 4-points amplitudes respectively with open or closed strings.
\end{abstract}

\keywords{soft-theorems; higher-spins; string perturbation theory.}

\bodymatter


\section{General introduction}
The study of the soft limit of scattering amplitudes is a well established topic in QFT since the 
pioneering works of Low~\cite{Low:1954kd} in quantum electrodynamics, and Weinberg~\cite{Weinberg1965}, Gross and Jackiw~\cite{GrossJackiw1968} in Einstein gravity. Recently, this subject has raised renewed attention 
due to the discovery~\cite{He:2014laa} of its connection with the B$^2$MS large diffeomorphisms invariance of general relativity for asymptotically flat spaces~\cite{Bondi:1962px,Barnich:2009se,Strominger:2013jfa,Cachazo:2014fwa,Hawking:2016msc,Kapec:2015vwa,Pasterski:2015tva,Strominger:2014pwa}.
Soft theorems have been derived for a wide class of QFTs and strings from the examination of scattering amplitudes at tree and loop level~\cite{Bern:2014vva,Bianchi:2014gla,Brandhuber:2015vhm,Luo:2015tat,Huang:2015sla,Volovich:2015yoa,Chen:2014cuc,Avery:2015iix,Schwab:2014sla,Low:2015ogb}. Concretely, if we consider color-ordered gluon amplitudes, the soft theorem relates the $n{+}1$ amplitude $\mathcal{A}_{n+1}(k_1,...,k_n;k_{n{+}1})$ to the $n$-points one $\mathcal{A}_n(k_1,...,k_n)$ through a differential operator  depending on the momentum and the polarization of the $(n{+}1)$-th external leg, when $k_{n{+}1}=\delta \hat{k}_{n{+}1}$ and $\delta\to 0$; in formulae
\beq
\mathcal{A}_{n+1}(k_1,...,k_n;k_{n+1})=\left(\frac{a_{n+1}k_1}{k_{n+1}k_1}{-}\frac{a_{n+1}k_n}{k_{n+1}k_n}{+}\frac{f_{n+1}{:}J_1}{k_{n+1}k_1}{-}\frac{f_{n+1}{:}J_n}{k_{n+1}k_n}\right)\mathcal{A}_n(k_1,...,k_n),
\label{softgluons}
\eeq
up to terms of order $\mathcal{O}(\delta^0)$. In Eq.~\eqref{softgluons} we introduced the field-strength $f_{\mu\nu}=a_\mu k_\nu{-}a_\nu k_\mu$, and the angular-momentum operator $J_{\mu\nu}$.
A stronger statement holds for gravity amplitudes: the soft behavior is ``universal" up to terms of order $\mathcal{O}(\delta)$, instead of order $\mathcal{O}(\delta^0)$ as for the Yang-Mills case
\begin{align}
&\mathcal{M}_{n+1}(k_1,...,k_n;k_{n+1})=\nonumber\\
&\qquad\sum_{i=1}^n\left( \frac{k_i h_{n+1}k_i}{k_ik_{n+1}}+\frac{k_ih_{n+1}J_ik_{n+1}}{k_ik_{n+1}}+\frac{k_{n+1}J_ih_{n+1}J_ik_{n+1}}{2k_ik_{n+1}} \right)\mathcal{M}_n(k_1,...,k_n).
\label{softgravs}
\end{align}
For open superstring and bosonic string theories, Eq.~\eqref{softgluons} describes the soft behavior of massless string 
amplitudes, notwithstanding the presence of an interaction term $\alpha^\prime F^3$ in the bosonic case, that in principle 
could spoil the validity of Eq.~\eqref{softgluons}. Closed superstring amplitudes satisfy Eq.~\eqref{softgravs}, while closed bosonic strings don't because at order $\mathcal{O}(\delta)$ there is a mixing between the graviton and the dilation, due to the interaction term $\alpha^\prime \varphi R^2$.
\section{Soft theorems in QFT and strings}
In this section we sketch in some detail the proof of Eq.~\eqref{softgluons} for color-ordered gluon amplitudes in QFT -- see Ref.~\cite{Bern:2014vva} for a more complete discussion, and 
compare this approach to the one relying on the OPE expansion presented in Ref.~\cite{Bianchi:2014gla}.
\subsection{Gluon amplitudes in QFT}
The singular soft behavior of the tree level color ordered amplitude $\mathcal{A}_{n{+}1}(k_1,...,k_n;k_s)$, when $k_s$ becomes soft, arises from the propagators in the channels $k_1{+}k_s$ and $k_n{+}k_s$, in formulae
\begin{align}
&\qquad\qquad\qquad\qquad\qquad\qquad\mathcal{A}_{n+1}(k_1,...,k_n;k_s)=\nonumber\\
&\frac{V_\mu(k_s,k_1)}{(k_s{+}k_1)^2}\mathcal{A}_n^\mu(k_1{+}k_s,...,k_n){+}\frac{V_\mu(k_n,k_s)}{(k_s{+}k_n)^2}\mathcal{A}_n^\mu(k_1,...,k_n{+}k_s){+}R_{n{+}1}(k_1,...,k_n;k_s).
\label{TreeAmp}
\end{align}
The Yang-Mills vertex for incoming momenta $k_s,k_i,k_I$, with $k_I{=}{-}k_i{-}k_s$, can be written as
\beq
V_\mu(k_s,k_i)=V_{\mu,i}^{(0)}+V_{\mu,i}^{(1)}=2a_i^\mu\,a_sk_i+2k_s^\mu\,a_sa_i-2a_s^\mu\,a_ik_s,
\label{YMexp}
\eeq
\begin{align}
V^{(0)}_{\mu,i}{=}V_\mu(0,k_i){=}2a_i^\mu\,a_sk_i\qquad\text{and}\qquad V_{\mu,i}^{(1)}{=}k_s\frac{\de}{\de k_s}V_\mu(0,k_i){=}2k_s^\mu\,a_sa_i{-}2a_s^\mu\,a_ik_s.
\end{align}
The expansion of the $n$-point amplitude $\mathcal{A}_n$ around $k_i$ yields 
\beq
\mathcal{A}^\mu_n(...,k_s{+}k_i,...){=}\mathcal{A}_{n,i}^{\mu(0)}{+}\mathcal{A}_{n,i}^{\mu(1)}{+}...{=}\mathcal{A}_n^\mu(...,k_i,...){+}k_s\frac{\de}{\de \hat{k}_i}\mathcal{A}_n^\mu(...,\hat{k_i},...)|_{\hat{k}_i=k_i}{+}...\,,
\eeq
Combining it with Eq.~\eqref{YMexp} produces the full soft expansion of Eq.~\eqref{TreeAmp}.
Imposing gauge invariance order by order in the soft parameter $\delta$, we can determine the leading term of the 
unknown contribution $R_{n{+}1}(k_1,...,k_n;k_s)$.
At leading order
\beq
\frac{V^{(0)}_{\mu,1}}{2k_sk_1}\mathcal{A}_{n,1}^{\mu(0)}{-}\frac{V^{(0)}_{\mu,n}}{2k_sk_n}\mathcal{A}_{n,n}^{\mu(0)}=\left( \frac{a_sk_1}{k_sk_1} - \frac{a_sk_n}{k_sk_n} \right)\mathcal{A}_n(k_1,...,k_n)\overset{a_s\to k_s}\longrightarrow0.
\eeq
At sub-leading order, it is possible to recognize a manifest gauge invariant contribution
\beq
\frac{V^{(1)}_{\mu,1}}{2k_sk_1}\mathcal{A}_{n,1}^{\mu(0)}-\frac{V^{(1)}_{\mu,n}}{2k_sk_n}\mathcal{A}_{n,n}^{\mu(0)}{=}\frac{f_s^{\mu\nu}a_{1[\nu}\de/\de a_1^{\mu]}}{2k_sk_1}\mathcal{A}_{n}^{(0)}{-}\frac{f_s^{\mu\nu} a_{n[\nu} \de/\de a_n^{\mu]} }{2k_sk_n}\mathcal{A}_{n}^{(0)}\overset{a_s\to k_s}\longrightarrow 0,
\eeq
which reconstructs the action of the spin part of the angular momentum, and
a non gauge invariant part
\begin{align}
&\frac{V^{(0)}_{\mu,1}}{2k_sk_1}\mathcal{A}_{n,1}^{\mu(1)}{-}\frac{V^{(0)}_{\mu,n}}{2k_sk_n}\mathcal{A}_{n,n}^{\mu(1)}{+}a_{s}\widetilde{R}_{n+1}(k_1,...,k_n;k_s{=}0)\nonumber\\
&\overset{a_s\to k_s}\longrightarrow k_s\frac{\de}{\de k_1}\mathcal{A}_{n}-k_s\frac{\de}{\de k_n}\mathcal{A}_{n}+k_s\widetilde{R}_{n+1}(k_s=0)=0.
\label{subgauge}
\end{align}
Solving Eq.~\eqref{subgauge} for $R_{n+1}(k_s{=}0)$ allows us to 
 recover the action of the orbital part of the angular-momentum.
\subsection{Gluon amplitudes in string theory}
\label{secstring}
The above analysis, performed for a QFT, can be reproduced as well by the OPE of vertex operators in string theory.
To compute and $n{+}1$ superstring gluon amplitude, we need two operators in the $q{=}{-}1$ super-ghost picture 
and $n{-}2$ vertices with $q{=}0$. Let us suppose the soft vertex operator to be integrated and in the $q{=}0$ super-ghost picture, and the adjacent operators to be in the $q{=}{-}1$ picture.
The leading order in $k_s$, when $z_s$ approaches $z_{s{+}1}$, is given by
\begin{align}
&\int^{z_{s{+1}1}} dz_s\,a_si\de X\,e^{ik_s X}(z_s)\,a_{s{+}1}\psi e^{-\varphi}e^{ik_{s{+}1}X}(z_{s{+}1}){\dots}\sim\nonumber\\
&\int^{z_{s{+1}1}} dz_s\,a_sk_{s{+}1}(z_s{-}z_{s{+}1})^{k_s k_{s{+}1}}{\dots}\approx \frac{a_sk_{s{+1}}}{k_sk_{s{+}1}}{\dots} 
\label{OPE}
\end{align}
Taking the sub-leading order in the soft expansion of Eq.~\eqref{OPE}, it happens that BRST invariance is responsible for 
reproducing the action of the orbital part of the angular momentum, as well as gauge invariance does in Eq.~\eqref{subgauge}~\cite{Bianchi:2014gla}.
The OPE of the manifestly BRST invariant sub-leading term $\frac{1}{2}\psi^\mu f_{\mu\nu}\psi^\nu\,e^{ikX}$ reproduces 
the action of the spin part of the soft operator.
The above analysis is remarkable because can be easily generalized when the OPE is performed between the soft operator and a massive higher-spin string belonging to the leading Regge trajectory.
\subsection{Soft behavior of disk scattering amplitudes with massive states}
In the open bosonic string, at the first massive level, we have only one BRST-invariant state given by the vertex operator
\beq
V_H{=}H_{\mu\nu}i\de X^\mu i\de X^\nu \,e^{ipX}
\eeq
with $p^\mu H_{\mu\nu}{=}0$, symmetric and traceless. Conversely, in the open superstring we have two different states interpolating in the $q{=}{-}1$ picture by
\beq
V_H{=}H_{\mu\nu}i\de X^\mu\psi^\nu\,e^{-\varphi}\,e^{ip X}\quad \text{and \quad }V_C{=}C_{\mu\nu\rho}\psi^\mu\psi^\nu\psi^\rho\,e^{-\varphi}\,e^{ipX},
\eeq
with $p^\mu C_{\mu\nu\rho}{=}0$ and completely antisymmetric.
In Ref.~\cite{Bianchi:2015yta} we have shown that string disk-amplitudes with the insertions of the aforementioned vertex operators satisfy Eq.~\eqref{softgluons}.
Following the steps outlined in Sec.~\ref{secstring}, it is straightforward to show that also amplitudes with the insertion 
of massive higher-spin strings like
\beq
V_{H^s}=H_{\mu_1\dots\mu_s} i\de X^{\mu_1}\dots i\de X^{\mu_{s-1}} \psi^{\mu_s}\,e^{-\varphi}\,e^{ipX}
\eeq
satisfy the soft theorem.
\section{Massive string amplitudes from Yang-Mills}
In Ref.~\cite{Mafra:2011nv} it is shown that open string disk-amplitudes can be expressed in terms of Yang-Mills 
tree-amplitudes covered by suitable generalized hypergeometric functions
\beq
\mathcal{A}_n^{ST}(1,...,n){=}\sum_{\sigma\in S_{n{-}3}} F(1,[2_\sigma,...,(n{-}2)_\sigma],n{-}1,n)\mathcal{A}_n^{YM}(1,[2_\sigma,...,(n{-}2)_\sigma],n{-}1,n).
\label{MSS}
\eeq
For $n=5$ Eq.~\eqref{MSS} yields
\beq
\mathcal{A}^{ST}_5(12345){=}F(1[23]45)\mathcal{A}^{YM}_5(1[23]45){+}F(1[32]45)\mathcal{A}^{YM}_5(1[32]45).
\label{5pt}
\eeq
As we suggested in Ref~\cite{Bianchi:2015yta}, it is possible to get 4-point amplitudes with massive states taking the residue at the massive pole in some 2-particle channels of the 5-points amplitude with massless strings
\beq
\lim_{s_{12}\to -1} (s_{12}{+}1)\mathcal{A}_n^{ST}(1,2,...,n){=}\sum_{S'}\mathcal{A}_3^{ST}(1,2,S')\otimes \mathcal{A}_{n{-}1}(S',3,...,n).
\eeq
\subsection{Dimensional reduction at $D=4$}
Before showing an explicit example, let us briefly review the content of the first massive level of the superstring theory in the bosonic sector. The first massive super-multiplet in $D=10$ yields a long multiplet of $\mathcal{N}=4$ in $D=4$
\beq
\{H_{MN},C_{MNP}\}\quad\rightarrow\quad\{H_{\mu\nu}, 8 \psi_\mu, 27 Z_\mu, 48 \chi, 42 \varphi\}.
\eeq
The bosonic degrees of freedom are related to the $10$-dimensional vertex operators as follows~\cite{Bianchi:2015yta}
\beq
H_{\mu\nu} \leftarrow H_{\mu\nu} \qquad
27 \,Z_\mu \leftarrow 6 \,H_{\mu,i}, 15 \,C_{\mu,ij}, 6\,C_{\mu\nu,i} \qquad
42\,\varphi \leftarrow 21 \,H_{ij}, 20 \,C_{ijk}, C_{\mu\nu\rho}.
\eeq 
Among these states, only three of them are $R$-symmetry singlets in $D=4$: the spin$-2$ $H_{\mu\nu}$, the pseudo-scalar $C_0{=}\varepsilon_{\mu\nu\rho\sigma} p^\sigma C^{\mu\nu\rho}/M$ and the scalar $H_0$ coming from the fact that the BRST-invariant string state $H_{MN}$ is traceless only in $D{=}10$, therefore after dimensional reduction we are left with the partial trace $H_{ij}{=}H_0\delta_{ij}/2$.
\subsection{Massive string amplitudes in $D=4$}
In $D=4$ there are only MHV or anti-MHV 5-points gluon amplitudes. 
For example, fixing the helicity of the gluons, Eq.~\eqref{5pt} yields
\beq
\mathcal{A}_5^{ST}(1^-2^-3^+4^+5^+){=}\frac{\lef 12 \re^3}{\lef 23 \re \lef 34 \re \lef 45 \re \lef 51 \re}F(1[23]45){+}\frac{\lef 12 \re^4}{\lef 13 \re \lef 32 \re \lef 24 \re \lef 45 \re \lef 51 \re} F(1[32]45).
\eeq
Taking the residue at the pole $s_{12}{+}1$, we get
\begin{align}
&\lim_{s_{12}\to {-}1}(s_{12}{+}1)\mathcal{A}_5^{ST}(1^-2^-3^+4^+5^+){=}\mathcal{B}(1{-}\alpha^\prime s, 1{-}\alpha^\prime t)\times\nonumber\\
&\frac{\lef 12 \re^3}{\lef 23 \re \lef 34 \re \lef 45 \re \lef 51 \re}[s_{23}s_{35}{+}(s_{34}{+}s_{35})s_{24}]{-}\frac{\lef 12 \re^4}{\lef 13 \re \lef 32 \re \lef 24 \re \lef 45 \re \lef 51 \re}s_{13}s_{24}.
\end{align}
It is remarkable to notice that a totally symmetric and traceless state, as $H_{\mu\nu}$, or any other spin$-s$ state $S_{\mu_1...\mu_s}$ couples only to vector bosons with opposite helicity, let's say $f_1^\pm$ and $f_2^\mp$. Conversely, a scalar state couples only to vector bosons with the same helicity $f_1^\pm$ and $f_2^\pm$.
This can be easily understood looking at the properties of the 't Hooft symbols which form a basis for the selfdual or anti-selfdual 
$4\times 4$ matrices
\begin{align}
(f_1^+)_{\mu\nu}(f_2^-)^{\nu\rho}&\propto \eta^a_{\mu\nu}\bar \eta^{b\nu\rho}=S^{ab\rho}_{\,\,\,\,\,\,\mu}\\
(f_1^+)_{\mu\nu}(f_2^+)^{\nu\rho}&\propto \eta^a_{\mu\nu} \eta^{b\nu\rho}=\delta^{ab}\delta^\rho_\mu+\varepsilon^{abc}\eta^{c\,\,\,\rho}_{\,\,\mu}.
\end{align}
This observation allows us to disentangle the contribution due to the polarization $H^{++}$ or $H^{--}$ from those due to the propagation of $H_0$ or $C_0$~\cite{Bianchi:2015yta}
\begin{align}
\mathcal{A}_5^{ST}(1^-2^-3^+4^+5^+) & \overset{s_{12}\to {-}1}\longrightarrow \frac{\lef 12 \re^2}{M} \times \mathcal{B}(1{-}\alpha^\prime s,1{-}\alpha^\prime t) \frac{M [35]}{\lef 34 \re \lef 45 \re}\qquad C_0/H_0\\
\mathcal{A}_5^{ST}(1^+2^+3^-4^-5^+) & \overset{s_{12}\to {-}1}\longrightarrow \frac{[ 12 ]^2}{M} \times \mathcal{B}(1{-}\alpha^\prime s,1{-}\alpha^\prime t) \frac{\lef 34 \re^3 [35]}{M^3 \lef 45 \re}\qquad C_0/H_0\\
\mathcal{A}_5^{ST}(1^+2^-3^-4^+5^+) & \overset{s_{12}\to{-}1}\longrightarrow M \times \mathcal{B}(1{-}\alpha^\prime s,1{-}\alpha^\prime t) \frac{\lef 13 \re^4 [35]}{M \lef 12 \re^2 \lef 34 \re \lef 45 \re}\qquad H^{++}/H^{--}.
\end{align}
Using $SO(3)$ little group transformations that leave unchanged the momentum of the massive particle $p=u\bar u{+}v\bar v$, it is straightforward to get the expression of the amplitude for all the other polarizations of the tensor $H$. 
Defining
\begin{align}
L_x\quad :\quad &u^\prime=\frac{1}{\sqrt{2}}(u+v)\quad\quad v^\prime=\frac{1}{\sqrt{2}}(-u+v)\\
L_y\quad :\quad &u^\prime=\frac{1}{\sqrt{2}}(u+iv)\quad\quad v^\prime=\frac{1}{\sqrt{2}}(iu+v),
\label{operators}
\end{align}
and noticing that $H^{++}{+}H^{- -}\overset{L_x{-}L_y}\longrightarrow H^{0}$, $H^{++}{-}H^{- -}\overset{L_x}\longrightarrow (H^-{-}H^+)/2$, and $H^{++}{-}H^{- -}\overset{L_y}\longrightarrow i(H^-{+}H^+)/2$, we get the full amplitude rotating the polarizations using linear combinations of the operators in Eq.~\eqref{operators}
\begin{align}
&\sum_hc_h\mathcal{A}(1^-2^+3^+H^h){=}\mathcal{B}(1-\alpha^\prime s,1-\alpha^\prime t)\nonumber\\
&\times\frac{[13]\lef 14 \re^2 \lef 15 \re^2}{M \lef 12 \re \lef 23 \re \lef 45 \re^2}\left(c_{++}\frac{\lef 14 \re^2}{\lef 15 \re^2}{-}4c_{+0}\frac{\lef 14 \re^2}{\lef 15 \re^2}{+}6c_{00}\frac{\lef 14 \re^2}{\lef 15 \re^2}{-}4c_{0-}\frac{\lef 14 \re^2}{\lef 15 \re^2}{+}c_{- -}\frac{\lef 14 \re^2}{\lef 15 \re^2}\right).
\end{align}
\section{Closed superstring amplitudes}
We conclude with some few comments about the soft behavior of closed string amplitudes with gravitons and massive states studied in detail in Ref.~\cite{Bianchi:2015lnw}.
Using the amplitudes computed in Ref.~\cite{Bianchi:2015yta}, and the KLT formula~\cite{KLT} 
\beq
\mathcal{M}_4(\mathcal{E}_1,\mathcal{E}_2,\mathcal{E}_3,\mathcal{K}_4{+}\mathcal{L}_4{+}\mathcal{U}_4){=}\sin\left( \pi \frac{\alpha^\prime t}{4} \right) \mathcal{A}_{4L}(1,2,3,H_4{+}C_4)\otimes \mathcal{A}_{4R}(1,3,2,H_4{+}C_4).
\eeq
We checked the correct soft behavior up to the sub-sub-leading order in $D=4$ using the spinor-helicity formalism for amplitudes involving gravitons, dilatons, and the massive states $\mathcal{K}$ and $\mathcal{H}$
\beq
\mathcal{M}_4(1^{-2},2^{+2},3^{+2},\mathcal{K}_4^{+4})\quad\mathcal{M}_4(1^{0},2^{0},3^{+2},\mathcal{K}_4^{+4})\quad\mathcal{M}_4(1^{0},2^{+2},3^{+2},\mathcal{H}_4^{+2}),
\eeq
when the graviton with momentum $k_3$ becomes soft.
While for bosonic closed string amplitudes we have found a discrepancy at sub-sub-leading order due to $\alpha^\prime$
dependent terms in agreement with Ref.~\cite{divecchiaetal}.
\section*{Acknowledgments}
We wish to thank A. Addazi, D. Consoli, P. Di Vecchia, Y. Huang, R. Marotta, F. Morales, L. Pieri, O. Schlotterer, S. Stieberger, G. Veneziano and C. Wen for many precious discussions and comments.


\begin{thebibliography}{9}
\bibitem{Low:1954kd} 
  F.~E.~Low,
  Phys.\ Rev.\  {\bf 96}, 1428 (1954).

\bibitem{Weinberg1965} 
  S.~Weinberg,
  Phys.\ Rev.\  {\bf 140}, B516 (1965).

\bibitem{GrossJackiw1968} 
  D.~J.~Gross and R.~Jackiw,
  Phys.\ Rev.\  {\bf 166}, 1287 (1968).

\bibitem{He:2014laa} 
  T.~He, V.~Lysov, P.~Mitra and A.~Strominger,
  JHEP {\bf 1505}, 151 (2015)
[\href{http://arXiv.org/abs/1401.7026}{arXiv:1401.7026 [hep-th]}].

\bibitem{Bondi:1962px} 
  H.~Bondi, M.~G.~J.~van der Burg and A.~W.~K.~Metzner,
  Proc.\ Roy.\ Soc.\ Lond.\ A {\bf 269}, 21 (1962).

\bibitem{Barnich:2009se} 
  G.~Barnich and C.~Troessaert,
  Phys.\ Rev.\ Lett.\  {\bf 105}, 111103 (2010)
  [\href{http://arXiv.org/abs/0909.2617}{arXiv:0909.2617 [gr-qc]}];
%
  PoS, 010 (2010)
  [Ann.\ U.\ Craiova Phys.\  {\bf 21}, S11 (2011)]
  [\href{http://arXiv.org/abs/1102.4632}{arXiv:1102.4632 [gr-qc]}];
  JHEP {\bf 1112}, 105 (2011)
  [\href{http://arXiv.org/abs/1106.0213}{arXiv:1106.0213 [hep-th]}].

\bibitem{Strominger:2013jfa} 
  A.~Strominger,
  JHEP {\bf 1407}, 152 (2014)
  [\href{http://arXiv.org/abs/1312.2229}{arXiv:1312.2229 [hep-th]}].

\bibitem{Cachazo:2014fwa} 
  F.~Cachazo and A.~Strominger,
  [\href{http://arXiv.org/abs/1404.4091}{arXiv:1404.4091 [hep-th]}].

\bibitem{Hawking:2016msc} 
  S.~W.~Hawking, M.~J.~Perry and A.~Strominger,
[\href{http://arXiv.org/abs/1601.00921}{arXiv:1601.00921 [hep-th]}].

\bibitem{Kapec:2015vwa} 
  D.~Kapec, V.~Lysov, S.~Pasterski and A.~Strominger,
[\href{http://arXiv.org/abs/1502.07644}{arXiv:1502.07644 [gr-qc]}].

\bibitem{Pasterski:2015tva} 
  S.~Pasterski, A.~Strominger and A.~Zhiboedov,
[\href{http://arXiv.org/abs/1502.06120}{arXiv:1502.06120 [hep-th]}].
  
\bibitem{Strominger:2014pwa} 
  A.~Strominger and A.~Zhiboedov,
[\href{http://arXiv.org/abs/1411.5745}{arXiv:1411.5745 [hep-th]}].

\bibitem{Bern:2014vva} 
  Z.~Bern, S.~Davies, P.~Di Vecchia and J.~Nohle,
  Phys.\ Rev.\ D {\bf 90}, no. 8, 084035 (2014)
  [\href{http://arXiv.org/abs/1406.6987}{arXiv:1406.6987 [hep-th]}].

\bibitem{Bianchi:2014gla} 
  M.~Bianchi, S.~He, Y.~t.~Huang and C.~Wen,
  Phys.\ Rev.\ D {\bf 92}, no. 6, 065022 (2015)
  [\href{http://arXiv.org/abs/1406.5155}{arXiv:1406.5155 [hep-th]}].

\bibitem{Brandhuber:2015vhm} 
  A.~Brandhuber, E.~Hughes, B.~Spence and G.~Travaglini,
[\href{http://arXiv.org/abs/1511.06716}{arXiv:1511.06716 [hep-th]}].

\bibitem{Luo:2015tat} 
  H.~Luo and C.~Wen,
[\href{http://arXiv.org/abs/1512.06801}{arXiv:1512.06801 [hep-th]}].

\bibitem{Huang:2015sla} 
  Y.~t.~Huang and C.~Wen,
  JHEP {\bf 1512}, 143 (2015)
[\href{http://arXiv.org/abs/1509.07840}{arXiv:1509.07840 [hep-th]}].

\bibitem{Volovich:2015yoa} 
  A.~Volovich, C.~Wen and M.~Zlotnikov,
  JHEP {\bf 1507}, 095 (2015)
[\href{http://arXiv.org/abs/1504.05559}{arXiv:1504.05559 [hep-th]}].

\bibitem{Chen:2014cuc} 
  W.~M.~Chen, Y.~t.~Huang and C.~Wen,
  JHEP {\bf 1503}, 150 (2015)
[\href{http://arXiv.org/abs/1412.1811}{arXiv:1412.1811 [hep-th]}];
  Phys.\ Rev.\ Lett.\  {\bf 115}, no. 2, 021603 (2015)
[\href{http://arXiv.org/abs/1412.1809}{arXiv:1412.1809 [hep-th]}].

\bibitem{Avery:2015iix} 
  S.~G.~Avery and B.~U.~W.~Schwab,
[\href{http://arXiv.org/abs/1512.02657}{arXiv:1512.02657 [hep-th]}];
[\href{http://arXiv.org/abs/1510.07038}{arXiv:1510.07038 [hep-th]}];
  Phys.\ Rev.\ D {\bf 93}, no. 2, 026003 (2016)
[\href{http://arXiv.org/abs/1506.05789}{arXiv:1506.05789 [hep-th]}].
  
\bibitem{Schwab:2014sla} 
  B.~U.~W.~Schwab,
  JHEP {\bf 1503}, 140 (2015)
[\href{http://arXiv.org/abs/1411.6661}{arXiv:1411.6661 [hep-th]}];
  JHEP {\bf 1408}, 062 (2014)
[\href{http://arXiv.org/abs/1406.4172}{arXiv:1406.4172 [hep-th]}].

\bibitem{Low:2015ogb} 
  I.~Low,
[\href{http://arXiv.org/abs/1512.01232}{arXiv:1512.01232 [hep-th]}].

\bibitem{Mafra:2011nv} 
  C.~R.~Mafra, O.~Schlotterer and S.~Stieberger,
  Nucl.\ Phys.\ B {\bf 873}, 419 (2013)
  [\href{http://arXiv.org/abs/1106.2645}{arXiv:1106.2645 [hep-th]}];
%
  Nucl.\ Phys.\ B {\bf 873}, 461 (2013)
  [\href{http://arXiv.org/abs/1106.2646}{arXiv:1106.2646 [hep-th]}].

\bibitem{Bianchi:2015yta} 
  M.~Bianchi and A.~L.~Guerrieri,
  JHEP {\bf 1509}, 164 (2015)
  [\href{http://arXiv.org/abs/1505.05854}{arXiv:1505.05854 [hep-th]}];
  A.~L.~Guerrieri,
  [\href{http://arXiv.org/abs/1507.08829}{arXiv:1507.08829 [hep-th]}].

\bibitem{Bianchi:2015lnw} 
  M.~Bianchi and A.~L.~Guerrieri,
  [\href{http://arXiv.org/abs/1512.00803}{arXiv:1512.00803 [hep-th]}].

\bibitem{KLT} 
  H.~Kawai, D.~C.~Lewellen and S.~H.~H.~Tye,
  Nucl.\ Phys.\ B {\bf 269}, 1 (1986).

\bibitem{divecchiaetal} 
  P.~Di Vecchia, R.~Marotta and M.~Mojaza,
  JHEP {\bf 1505}, 137 (2015)
  [\href{http://arXiv.org/abs/1502.05258}{arXiv:1502.05258 [hep-th]}];
  P.~Di Vecchia, R.~Marotta and M.~Mojaza,
  [\href{http://arXiv.org/abs/1511.04921}{arXiv:1511.04921 [hep-th]}];
  P.~Di Vecchia, R.~Marotta, M.~Mojaza and J.~Nohle;
  [\href{http://arXiv.org/abs/1512.03316}{arXiv:1512.03316 [hep-th]}].

\end{thebibliography}
\end{document}